\documentclass[review]{elsarticle}
\usepackage{hyperref}

\usepackage{url}

\usepackage{soul}
\usepackage{subcaption}
\usepackage{longtable}
\usepackage{makecell}
\usepackage{hyperref}
\usepackage{enumitem}
\newlist{Properties}{enumerate}{2}
\setlist[Properties]{label=\textbf{Property \arabic*.},itemindent=*}
\usepackage{mathtools}
\usepackage[thmmarks, amsmath]{ntheorem}%

\theoremheaderfont{\bfseries\upshape}
\newtheorem{defi}{Definition}

\usepackage{amsmath,amssymb,amsfonts}
\usepackage{algorithm,algorithmicx,algpseudocode}
\algrenewcommand\algorithmicrequire{\textbf{Input:}}
\algrenewcommand\algorithmicensure{\textbf{Output:}}
\algdef{SE}[DOWHILE]{Do}{doWhile}{\algorithmicdo}[1]{\algorithmicwhile\ #1}%
\usepackage[font=footnotesize,labelfont=bf]{subcaption}
\usepackage{graphicx}
\usepackage{xcolor}
\usepackage[utf8]{inputenc}
\usepackage{csquotes}
\usepackage[font=footnotesize,labelfont=bf]{caption}
\makeatletter
\newcommand{\ssymbol}[1]{^{\@fnsymbol{#1}}}
\makeatother

\begin{document}

\begin{frontmatter}

\title{Towards Situational Aware Cyber-Physical Systems: \\ A Security-Enhancing Use Case of Blockchain-based Digital Twins}

\author[First]{Sabah Suhail\corref{cor1}\fnref{label2}}
\ead{sabah.suhail@wu.ac.at}
\address[First]{Research Institute for Cryptoeconomics, Information Systems and Operations Management, Vienna University of Economics and Business, Vienna, Austria.}

\author[Second]{Saif Ur Rehman Malik}
\ead{saif.rehmanmalik@cyber.ee}
\address[Second]{Cybernetica AS, Tallinn, Estonia.}

\author[Third]{Raja Jurdak}
\ead{r.jurdak@qut.edu.au}
\address[Third]{Queensland University of Technology, Brisbane, Australia.}

\author[Fourth]{Rasheed Hussain}
\ead{rasheed.hussain@bristol.ac.uk}
\address[Fourth]{Bristol Digital Futures Institute (BDFI) and Smart Internet Lab, University of Bristol, Bristol, UK.}

\author[Fifth]{Raimundas Matulevičius}
\ead{raimundas.matulevicius@ut.ee}
\address[Fifth]{University of Tartu, Tartu, Estonia.}

\author[First,Sixth]{Davor Svetinovic}
\ead{davor.svetinovic@wu.ac.at}
\address[Sixth]{Center for Cyber–Physical Systems, Electrical Engineering and Computer Science, Khalifa University of Science and Technology, Abu Dhabi, UAE.}

\begin{abstract}
The complexity of cyberattacks in Cyber-Physical Systems (CPSs) calls for a mechanism that can evaluate critical infrastructures' operational behaviour and security without affecting the operation of live systems. In this regard, Digital Twins (DTs) provide actionable insights through monitoring, simulating, predicting, and optimizing the state of CPSs. Through the use cases, including system testing and training, detecting system misconfigurations, and security testing, DTs strengthen the security of CPSs throughout the product lifecycle. However, such benefits of DTs depend on an assumption about data integrity and security. Data trustworthiness becomes more critical while integrating multiple components among different DTs owned by various stakeholders to provide an aggregated view of the complex physical system. This article envisions a blockchain-based DT framework as Trusted Twins for Securing Cyber-Physical Systems (TTS-CPS). With the automotive industry as a CPS use case, we demonstrate the viability of the TTS-CPS framework through a proof of concept. To utilize reliable system specification data for building the process knowledge of DTs, we ensure the trustworthiness of data-generating sources through Integrity Checking Mechanisms (ICMs). Additionally, Safety and Security (S\&S) rules evaluated during simulation are stored and retrieved from the blockchain, thereby establishing more understanding and confidence in the decisions made by the underlying systems. Finally, we perform formal verification of the TTS-CPS.
\end{abstract}


\end{frontmatter}

\section{Introduction}\label{introduction}
Characterized by computation, networking, and physical components, Cyber-Physical Systems (CPSs) interface the digital and physical  worlds~\cite{baheti2011cyber} to enable the realization of the Industry 4.0 vision~\cite{suhail2021securing}. Due to the fact that Industrial Control System (ICS)-a subset of CPSs have a direct impact on the environment they operate in, ensuring that such systems meet specific security and safety requirements is paramount~\cite{Eckhart2019}. Several seminal examples of ICS-tailored malware have demonstrated~{\it how severe the consequences of these incidents can be}~\cite{Eckhart2019, suhail2021blockchainbased}. For instance, the cyber attack on the Ukrainian power grid in 2015 (BlackEnergy3)~\cite{khan2016threat} and a follow-up attack in 2016 (Industroyer)~\cite{Industroyer} disconnected several substations, causing a power outage. Similarly, Stuxnet~\cite{langner2011stuxnet} targeting Iran uranium enrichment plant and Triton~\cite{miller2019triton} targeting a petrochemical plant in Saudi Arabia are among the most prominent examples of cyber espionage. By exploiting loopholes in the system infrastructure, the attackers gain a foothold and launch covert attacks or Advanced Persistent Threats (APTs)~\cite{suhail2021securing}. Consequently, such attacks degrade the overall system performance, cause significant economic loss, and pose human safety risks. Therefore, the specifically designed ICS-tailored covert attacks require a solution that, without obstructing the ongoing operations, can monitor and analyze the physical process to detect security loopholes in the CPS at an early stage (for instance, design phase), thereby reducing incident response time~\cite{suhail2021securing}.

Digital Twins (DTs) are considered one such solution.
Being the virtual replicas of their physical counterparts, DTs share the expected operational behaviour of the underlying systems~\cite{suhail2021securing} and provide a sustainable strategy for analyzing, monitoring, and predicting the behaviour of a system~\cite{corallo2021shop}. To do so, DTs collect data from multiple sources, such as installed sensors and actuators at the factory floor, historical production data derived from product lifecycle data, and domain knowledge. Following a closed feedback loop, DT inspects for data inconsistencies between the physical entity and virtual entity and feed back the simulation data to the physical entity to adopt better calibration and testing strategies~\cite{suhail2021trustworthy}. To fully harness the features of DT, the data used by the DT needs to be trustworthy and secure. For instance, reasoning about the current state of a data object entails a complete lineage of processes chain~\cite{suhail2019orchestrating}. Additionally, the requirement of data trustworthiness becomes more critical as it may affect the next system generation where DT data can be used as historical data. Unreliable data used as historical data in guiding future system iterations can lead to significant deviations from the system's desired behaviour. In this regard, leveraging blockchain technology allows industries to manage data on a distributed ledger while ensuring trusted DT data coordination~\cite{suhail2021blockchainbased, shen2021secure}. A provenance-enabled blockchain-based DT assure the traceability and solidity of data, thereby establishing more confidence in the decisions made by the underlying systems~\cite{suhail2021securing}. Thus, combining blockchain and DT can reshape the industry such that blockchain ensures secure data management and DTs use reliable data as input to extract actionable insights~\cite{suhail2021trustworthy}.

\begin{table}[ht!]
\centering
\caption{Existing works on specification-based digital twins: A summary.} \label{tab:survey}
\begin{tabular}{p{0.5cm} p{0.5cm} l p{6.50cm}}
\hline
\textbf{Year} & \textbf{Ref.} & {\thead{\textbf{Framework} /\\\textbf{Proof Of Concept (POC)}}} & \textbf{Objective of DTs}  \\ [0.5ex]
\hline
\hline
2018 & \cite{eckhart2018towards} & POC  & Detect security and safety rule violations 
\\ 
2018 & \cite{eckhart2018specification} & POC  & State-based intrusion detection
\\
 2019 &\cite{eckhart2019enhancing} & Framework & Risk assessment, Incident handling  
 \\
2019 & \cite{dietz2019distributed}  & Framework & Secure data sharing
 \\
2020 & \cite{dietz2020unleashing} & Framework & ICS security 
\\
2020 & \cite{dietz2020integrating} & POC & Incident analysis
\\
2021 & \cite{patel2021real}& POC & Attack detection and localization
\\ 
2021&\cite{vielberth2021digital} & POC & Utilizing cyber ranges for training security analysts
\\
\hline
\end{tabular}
\end{table}

In the information security domain, the concept of building the process knowledge of DTs is primarily based on either utilizing (i) system specification data to model the physical counterpart based on engineering artifacts~\cite{eckhart2018towards, eckhart2018specification, dietz2020integrating}, or (ii) Machine learning methods to learn security-related aspects based on sensor data~\cite{groshev2021toward, damjanovic2018digital} without obtaining process knowledge from DTs~\cite{suhail2021securing}. Focusing on specification-based DTs, present works do not consider whether data sources are trustworthy, which is critical to ensure input data quality. Similarly, the need for data storage is also not given due attention. Thus, trustworthy process knowledge for DTs is required to address the limitation in current knowledge.

\begin{table}[ht!]
 \renewcommand*{\arraystretch}{1.3}
  \centering
\caption{Notations and their explanation.} \label{tab:notations}
  \begin{tabular}{cccc }
  \hline
\textbf{Notation} & \textbf{Explanation} & \textbf{Notation} & \textbf{Explanation}   \\ [0.5ex]
 \hline 
 \hline
$S_{ID}$ & Sensor ID & $D_{s,t}$ & Timestamped sensor data \\
$A_{ID}$ & Asset ID & $A_\delta$ & Asset capacity  \\
$A^c_\omega$ & Asset current status  & $A^h_\omega$ & Asset history status \\
$\chi$ & Configuration settings & $D_P$ & Provenance data \\
$R_{ID}$ & Rule ID & $R_D$ & Rule description  \\
$E_{ID}$ & Entity ID & $P_{ID}$ & Process ID \\
$\tau$ & Threshold & $\epsilon$ & Process-specific settings\\
$\mathcal{R}$ & Set of rules &  $T_d$ &  task duration \\
$d$ & Delay & $V$ & Velocity\\
\hline
\end{tabular}
\end{table}

We focus on the existing research work covering the initial development phase of DTs, which consists of obtaining process knowledge through system specification approaches. Most importantly, we narrow down the existing works based on the objective of using DT, i.e., a situational awareness enabler. Table~\ref{tab:survey} summarizes relevant works on specification-based DTs. Moreover, Table~\ref{tab:survey} also shows that in recent years there has been an increased interest in using DTs to secure CPSs. The specification-based approaches (summarized in Table~\ref{tab:survey}) utilize technical, topological, and control artefacts of the underlying system that are maintained throughout the system engineering process~\cite{suhail2021securing}. There exist many works on blockchain-based DTs.~\cite{suhail2021blockchainbased} provides a comprehensive review of the design solutions for blockchain-based DTs in the industrial domain. However, the existing blockchain-based schemes lack the details on how the DTs are constructed and how DT security operation modes can be used to secure CPSs. Furthermore, they vary in terms of utilizing DTs and are therefore beyond the scope of this paper.~\cite{dietz2019distributed} proposed only a theoretical blockchain-based framework while neglecting the prototypical implementation.

Compared to the existing works, the proposed TTS-CPS scheme ensures that the generation of a virtual environment through system specification data is based on trustworthy data-generating sources owing to Integrity Checking Mechanisms (ICMs). Additionally, integrating blockchain makes the simulation environment reliable. In this work, we investigate the significance of integrating blockchain-based DTs in the CPS. We focus on specification-based DTs. More precisely, we propose an blockchain-based DTs framework for supporting security-enhancing use case of DTs in the CPS. 
Our main contributions can be summarized as follows:
\begin{itemize}
\item To support a situational-aware environment, we propose a blockchain-based DT framework as Trusted Twins for Securing Cyber-Physical Systems (TTS-CPS). By leveraging blockchain with DTs, we can track the accountable entity for adding or updating the Safety and Security (S\&S) rules and ensure the trustworthiness of data generating sources through ICMs.
\item Through a prototypical implementation supporting the simulated network topology, Human Machine Interfaces (HMIs), Programmable Logic Controllers (PLCs), and physical devices (e.g., robotic arm, motor), we demonstrate the feasibility of the TTS-CPS framework for an assembly line in the automotive industry. 
\item We perform formal verification of the TTS-CPS. 
\end{itemize}

The rest of the paper is organized as follows. Table~\ref{tab:notations} lists all the acronyms used in the paper. Section~\ref{framework} outlines the proposed framework for the CPS. Section~\ref{simulation} presents the evaluation results and discusses the formal modelling and verification of the proposed approach. Finally, Section~\ref{conclusion} concludes the paper with an outlook on future research directions.

\begin{figure}[ht!]
\centerline{\includegraphics[width=3.5in]{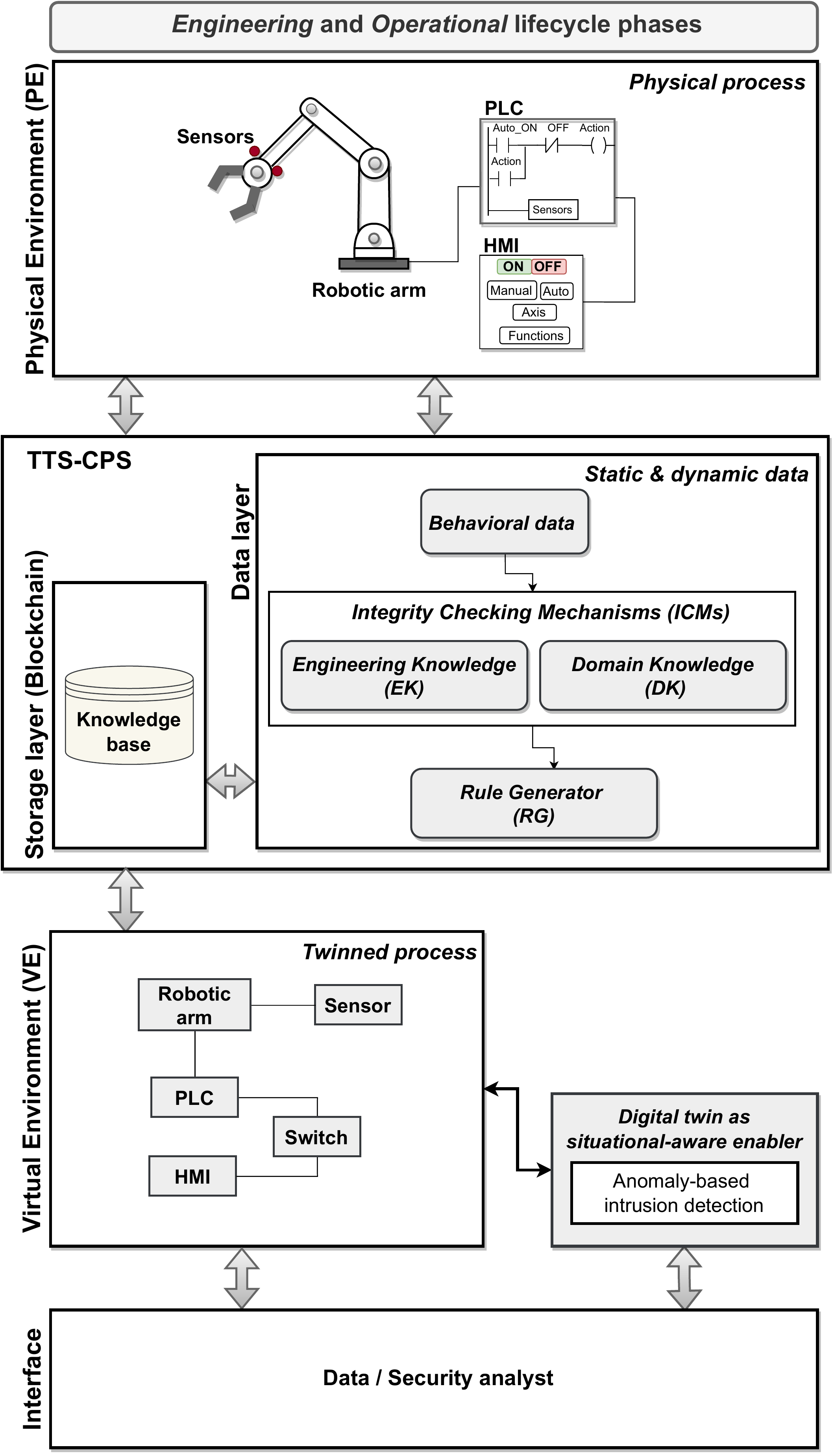}}
\caption{A high-level framework of TTS-CPS.}
\label{fig:framework1}
\end{figure}

\section{Framework Design: Trusted Twins for Securing Cyber-Physical Systems (TTS-CPS)}\label{framework}
This section describes the design of a blockchain-based DT framework called Trusted Twins for Securing Cyber-Physical Systems (TTS-CPS). We first present an overview of the TTS-CPS framework in Section~\ref{overview}. Then we provide a detailed discussion on the main components of the TTS-CPS, including comparison with existing frameworks in Section~\ref{comparison}, ICMs in Section~\ref{ICMs}, digital-physical mapping in Section~\ref{Digital-Physical}, and blockchain-based DTs in Section~\ref{BC_DTs}. 

\subsection{Overview of the proposed framework} \label{overview}
Fig.~\ref{fig:framework1} illustrates the high-level framework of the TTS-CPS. Firstly, the~{\it Physical Environment} (PE) showcases the physical process. Secondly, the~{\it Virtual Environment} (VE) showcases the twinned process. The physical process and its clone counterparts span engineering to operational lifecycle phases. Thirdly, the~{\it data layer} connects the physical and virtual environment by utilizing behavioural data. The behavioural data can be either static data (such as the range of sensor data) or dynamic data (such as real-time sensor data). The process knowledge of the DT is acquired from the specification-based artefacts. Fourthly, the~{\it storage layer} stores data in the knowledge base, which other layers can utilize. 
Finally, the~{\it interface} allows data/security analysts to repeat and reproduce simulations based on user-specified parameters and feedback during repetitive testing. Generally, the role of data/security analysts is (i) to utilize the controlled, supportive virtual environment during replication (record and replay events~\cite{eckhart2019enhancing}) and simulation (trial and error) modes, and (ii) to spot deviations from a defined or learned baseline to alert system. Among various security-enhancing use cases of DTs, including security testing, incident response, automated security, etc.~\cite{suhail2021securing}, we are using DTs as an anomaly detection tool. More specifically, the objective of DT is to identify~\emph{data inconsistencies} between PE and VE. 

\begin{figure}[ht!]
\centerline{\includegraphics[width=4.5in]{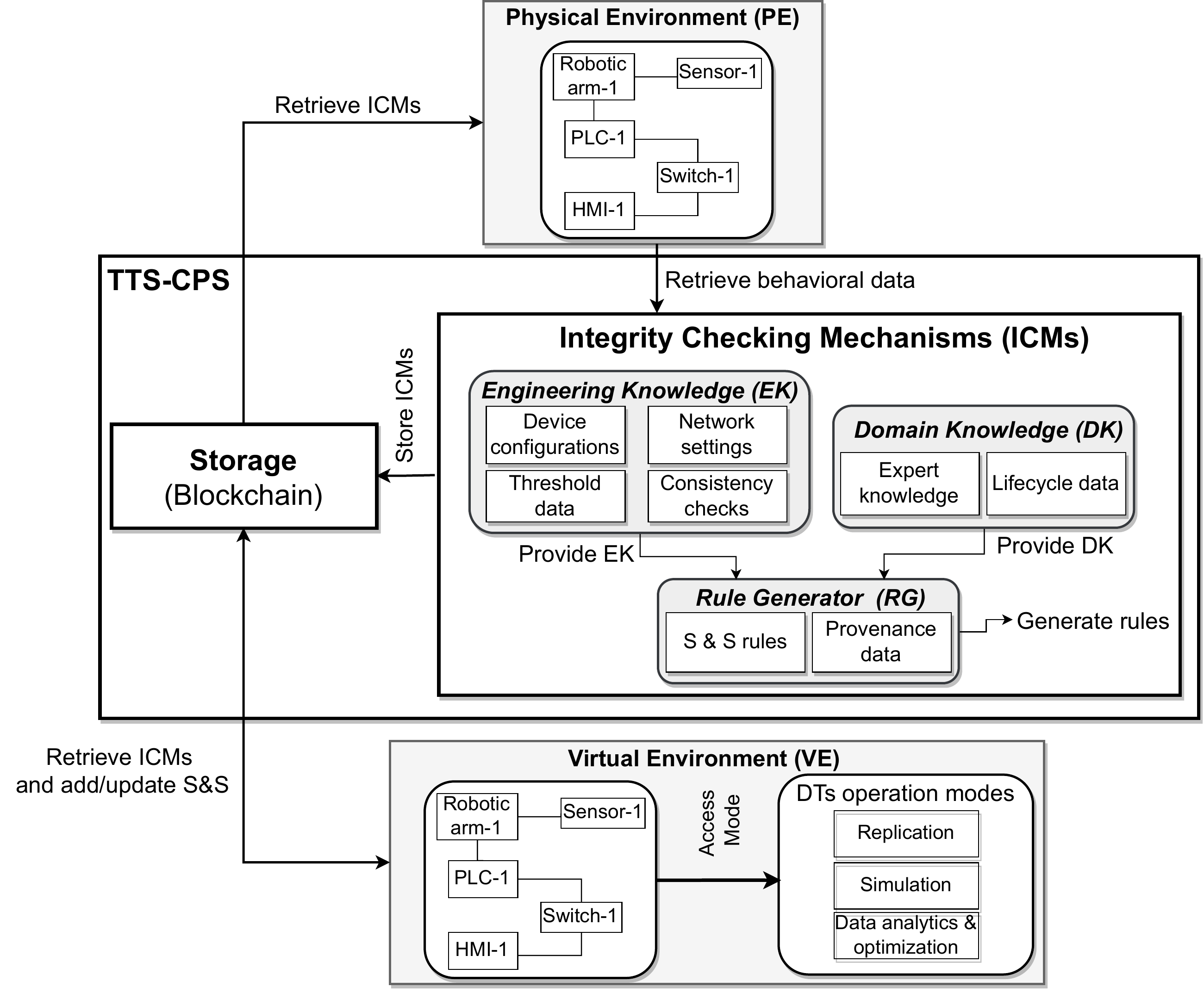}}
\caption{Blockchain-based digital twins: a secure CPS framework.}
\label{fig:framework2}
\end{figure}

To further elaborate the connection between the components of the proposed framework, we refer to Fig.~\ref{fig:framework2}. The framework comprises three key components, including (i) ICMs, (ii) PE and VE, and (iii) data storage (blockchain-based DTs). The~{\it ICMs} provide design specifications of the CPS necessary to generate the network setup of the virtual environment in addition to implicit S\&S rules. Based on ICMs components including Engineering Knowledge (EK), Domain Knowledge (DK), and Rule Generator (RG), a digital-physical mapping between~{\it PE} and its clone counterpart~{\it VE} to ensure data consistency between the two spaces is performed. The DT operation modes (simulation, replication, and data analytics) that support monitoring, behavioural analysis, and replaying of CPS events are part of the VE component. The~{\it blockchain ledger} enforces secure data management by storing data and recalling data or events.

In the following, we compare the proposed TTS-CPS framework with existing solutions closely aligned with our scope, i.e., securing CPS and using specification-based process knowledge for DTs.

\subsection{Comparison with existing frameworks} \label{comparison}
The DT paradigm qualifies for different enterprises to various implementation degrees~\cite{dietz2020digital} mainly due to (i) application-specific building blocks (such as assets, process knowledge, data sources) of DTs, (ii) objective of using DTs (such as cyber-situational awareness, predictive maintenance, resource optimization), and (iii) the level of details and granularity.
Considering the objective of using DTs to secure CPS, existing solutions such as CPS Twinning framework(~\cite{eckhart2019enhancing, eckhart2021digital}) utilize system specification data (i.e., EK and DK) to model the physical counterpart. However, these frameworks lack the concept of trusted twins, i.e., trustworthy process
knowledge for DTs due to the absence of blockchain-based storage in addition to the approach of leveraging system specification data as ICMs. 
TTS-CPS (adapted from ~\cite{eckhart2019enhancing, eckhart2021digital}) also uses specification-based process knowledge for DTs. However, we put emphasis on data trustworthiness and security which is the critical parameter in DT-based CPS security. Therefore, in addition to EK and DK, we have considered additional components such as RG and blockchain-based data storage. More specifically, we have used EK, DK, and RG (together as ICMs) to establish trusted twins.

Note that for TTS-CPS, we have mainly considered critical components to establish trusted twins. However, there could be additional components depending on the objectives of using DTs or application-specific requirements. For instance, with the similar objective of anomaly detection,~\cite{suhail2021trustworthy} envisioned a blockchain-based DT framework for the Industrial Internet of Things (IIoT). The main components of their proposed framework include (i) data wrangler (responsible for data conversion and data cleaning to transform the data into a unified form), (ii) data fusion (responsible for accumulating data from multiple DTs to cross-validate the observations to increase data consistency and trustworthiness), and (iii) data synchronization (responsible for digital-physical mapping and checking for data inconsistencies between physical and virtual space).
In this paper, we focus on the TTS-CPS components that promote trusted twins, and we reflect on the need for additional components, such as those in~\cite{suhail2021trustworthy}, for future research in Section~\ref{conclusion}.

\subsection{Integrity Checking Mechanisms (ICMs)} \label{ICMs}
The data layer mainly consists of ICMs. In the following, we discuss sub-components of the ICMs in detail. The~{\it quality} of the data matters at par with~{\it quantity} for ensuring precise predictions and decision-making in critical infrastructures operating under the presence of a complex threat landscape and high volume and variety of big data~\cite{suhail2021blockchainbased}. Although stringent security guarantees are inherited from the blockchain, ensuring the~\emph{trustworthiness} of data-generating sources is equally essential for critical CPS~\cite{suhail2021securing}. Since DTs act as the input data sources, DTs need to be built on reliable data. Therefore, a tri-fold ICMs is defined by the TTS-CPS framework. Firstly,~{\it Engineering Knowledge} (EK) provides the design specifications of the underlying CPS infrastructure to identify known devices and connections, authorized addressing and routing information, and expected behaviour of the process. Moreover, Calibration and Verification (C\&V) identify the root cause of aberrant device behaviour or erroneous data before data collection. Secondly,~{\it Domain Knowledge} (DK) provides domain-specific knowledge and system history data to predict equipment fault. Thirdly,~{\it Rule Generator} (RG) cross-validate the device data with overlapping fields of view by comparing pre-defined device performance parameters, S\&S rules, and provenance data.To give an example, RG takes input from EK and DK to generate S\&S rules while taking into account the provenance data to obtain the trustworthiness of data.
To sum up, ICMs provide reliable system specification data for building the process knowledge of DT at design phase. Based on specification-based process knowledge, DT acts as an anomaly detection tool. Additionally, the virtual network setup retrieves data from the blockchain, which establishes more understanding and confidence in the decisions made by the underlying systems.

\subsubsection{Engineering Knowledge (EK)}\label{subsec:EK}
Engineering Knowledge (EK) provides the design specifications of the technical, topological, and control artefacts at device-level, network-level, and system-level that are essential to generating the virtual environment's network setup. The device-level information includes construction details (e,g., name, type, make, model), functional details (e.g., operating conditions in terms of standards), and configuration details (e.g., IP and MAC address, I/O channels, control logic). The network-level information includes topology and communication path through logical connections and endpoints. The system-level information includes relationships among components/processes concerning data aggregation, conditional rules, and constraints. Each asset (including sensors) can be identified with asset ID ($A_{ID}$) and sensor ID ($S_{ID}$), whereas the associated configuration settings can be represented as $\chi$. 

Sensors are notoriously prone to calibration errors and arguably explain the root cause of aberrant behaviour or erroneous data. Ignoring such minor variations in the system behaviour may collapse the whole system or may incur drastic effects on the system's long-term behaviour. Calibration errors manifest when sensors report values that are offset from the ground truth. Given that factory calibration conditions may not necessarily be relevant to physical environment needs and sensors wear out over time, calibration is necessary to increase the sensor's accuracy and resiliency against random errors~\cite{ni2009sensor}. Therefore, to ensure data quality before data collection, sensors and actuators must be calibrated to ensure the measurement accuracy according to a known standard and verified to ensure the correct operation according to operating specifications. Exercising C\&V operations of Internet of Things (IoT) sensors must be carried out periodically for ageing management and fault diagnosis.

Each type-k sensor $s(k)$ $\in$ $\mathcal{K}$ where $\mathcal{K}= \{1, \ldots,k, \ldots, K\}$ collects timestamped sensor data $D_{s(k),t}$. To ensure that $D_{s(k),t}$ are within the predefined bounds, the following Consistency Check (CC) is performed:
\begin{equation}
\tau^\mathbf{min}_{s(k)} \leq D_{s(k),t} \leq \tau^\mathbf{max}_{s(k)},  \qquad \forall s(k) \in \mathcal{K},
\end{equation}
where both $\tau^\mathbf{min}_{s(k)}$ and $\tau^\mathbf{max}_{s(k)}$ $\in$ $\tau$ define the lower and upper bounds respectively.

\subsubsection{Domain Knowledge (DK)}\label{subsec:DK}
Domain Knowledge (DK) provides domain-specific knowledge from experts in various fields such as engineers (electrical, mechanical, instrumentation and control), supply chain participating entities, security professionals in Security Operations Center (SOC), etc. Moreover, it also includes asset historical data ($A^h_{\omega}$) derived from lifecycle data (such as design and development; operation and maintenance; and decommissioning). Once generated, DK can be used as a reference by different organizations and tailored accordingly to meet their specific needs.

\subsubsection{Rule Generator (RG)}\label{subsec:rules}
Rule Generator (RG) takes input from EK and DK to generate S\&S rules while taking into account the provenance data ($D_{P}$). Depending on the underlying CPS infrastructure, S\&S rules can be based on threshold data (upper $\tau^\mathbf{max}_{s(k)}$ and lower $\tau^\mathbf{min}_{s(k)}$ bounds), consistency checks (pre-defined performance parameters), and constraints (data accessibility and auditability)~\cite{suhail2021securing}. For instance, trends/patterns (e.g., heat or vibration), consistency checks (e.g., conveyor belt speed-variable), conditional limits for device data (e.g., minimum and maximum temperature), access control (e.g., authentication and authorization based on roles and access levels), etc. 

To detect the presence of malicious or accidental disruptions, the system needs to respond effectively by invoking the appropriate defence mechanisms, whereas the inability to impose such strategy results in long-term loss. Therefore, to detect misconfigurations and malicious activities, S\&S rules must be integrated into a CPS. S\&S rules monitor and analyze the device or process in the virtual environment and learn about the presence of an attack or abnormal behaviour by collecting data over such events. Based on the incident data, the derived patterns as S\&S rules can be formulated, tested, and transmitted to the physical environment. Thus, monitoring the physical system state through time, outliers, and changes can uncover anomalies or malicious activities by enabling the detection of possible S\&S rules violations in terms of deviations or patterns from a defined benign behaviour.

Introducing S\&S rules at DT's design and development phase can lower security and incident-response costs. Moreover, it makes later lifecycle phases (such as operation and maintenance; and decommissioning) less prone to errors and incidents~\cite{dietz2020unleashing}. To do so, S\&S rules can be defined at device-level or process-level. For instance, through EK defining a safe state based on the regular operation of the device, verifying the specific service provided by the process, deriving a white-list from network-level monitoring based on the authorized addressing and routing information, detecting unknown devices, identifying unidentified connections, determining abnormal changes in the control logic, enforcing fine- coarse- grained policies and constraints, etc.~\cite{suhail2021blockchainbased}. Similarly, rules can also be extracted from thresholds or consistency checks defined during the calibration phase and equipment history data from product lifecycle data. 

While reasoning about the current state and the chained actions on a data object (such as \textit{who}, \textit{when}, \textit{where}, \textit{why}, and \textit{how}), provenance data ($D_{P}$) can aid in generating device-, network-, and system-level rules. $D_{P}$ is a complete process lifecycle along with environment settings, input parameters, action and events performed on data~\cite{ZAFAR201750} and can be constructed as suggested in~\cite{suhail2019orchestrating}. The key role of $D_{P}$ is to enforce traceability in the CPSs while traversing through the process and can be stored on the blockchain to reconstruct the process chain on demand. $D_{P}$ can be reconstructed based on the data provided by EK, C\&V, and DK. Additionally, it can also serve as a basis for implicit security rules. For instance, $D_{P}$ under a process-specific settings ($P_{ID}$) can identify {\it who} is the accountable entity $E_{ID}$ generating or updating rule $R_{ID}$ defined as $R_{D}$ for $A_{ID}$ or $S_{ID}$.

\begin{equation}
P^{\epsilon}_{ID} = \{ R_{ID}, R_{D}, E_{ID}, A_{ID}, S_{ID} \}, \qquad \forall R \in \mathcal{R}.
\end{equation}

Similarly, $D_{P}$ can be derived from EK, C\&V, and DK to construct rules as follows. 
From EK, {\it what} are the configuration settings ($\chi$) defined for $S_{ID}$ affixed to $A_{ID}$?
\begin{equation}
D^{EK}_{P}=\{ A_{ID}, S_{ID}, \chi \}.
\end{equation}

From C\&V, {\it which} threshold settings ($\tau$) are optimal for $A_{ID}$ based on $A^{c}_{\omega}$ (i.e.,  on/off, run rate, speed) and $D_{s,t}$ from $S_{ID}$?
\begin{equation}
D^{C\&V}_{P}= \{ A_{ID}, A^{c}_{\omega}, S_{ID}, D_{s,t}, \tau \}.
\end{equation}

From DK, {\it how} $A_{ID}$ behaved under $\chi$?
\begin{equation}
D^{DK}_{P}= \{ A_{ID}, A^{h}_{\omega}, \chi \}.
\end{equation}

Depending on the severity of the cyber situation, S\&S rules needs to be generated or updated to make the system respond effectively and avoid long-term loss.

\begin{algorithm}[!ht]
\caption{Rules}\label{algo:rules}
\begin{algorithmic}[1]
\Require $E_{ID}$, $R_D$
\Ensure $\mathcal{R}$ in S\&S rules
\State Check access control rights of $E_{ID}$ 
\If{($R$ != $\mathcal{R}$)}
\State Assign $R_{ID}$
\State $R_{ID}$ $\leftarrow$ $R_D$
\State Associate $R_{ID}$ with $A_{ID}$ and/or $S_{ID}$
\State Store $R_{ID}$ at blockchain
\Else
\State Update $R_D$ of $R_{ID}$ 
\State Store $R_{ID}$ at blockchain
\EndIf
\end{algorithmic}
\end{algorithm}

Algorithm~\ref{algo:rules} illustrates the steps of generating or updating rules. Based on access control model such as Role-based Access Control (RBAC)~\cite{sandhu2000nist}, first check whether the entity ID ($E_{ID}$) has rights to generate or update rules. In case of a new rule $R$, a unique ID ($R_{ID}$) is assigned. Then $R$ is associated with respective asset or sensor for which it has been defined. In case of existing rule, rule description ($R_D$) is updated. Finally, the newly generated or updated rules are stored in the blockchain.

 \begin{figure}
        \centering
        \begin{subfigure}[b]{0.275\textwidth}
            \centering
        \includegraphics[width=\textwidth]{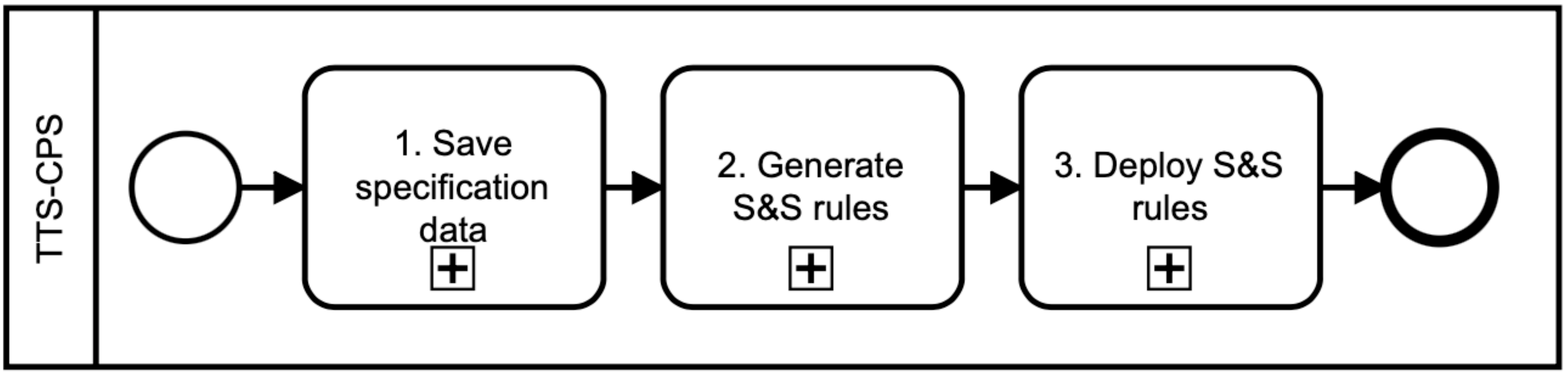} 
            \caption[Legend]%
            {{\small Legend}}    
            \label{fig:legend}
        \end{subfigure}
        \begin{subfigure}[b]{0.375\textwidth}  
            \centering 
         \includegraphics[width=\textwidth]{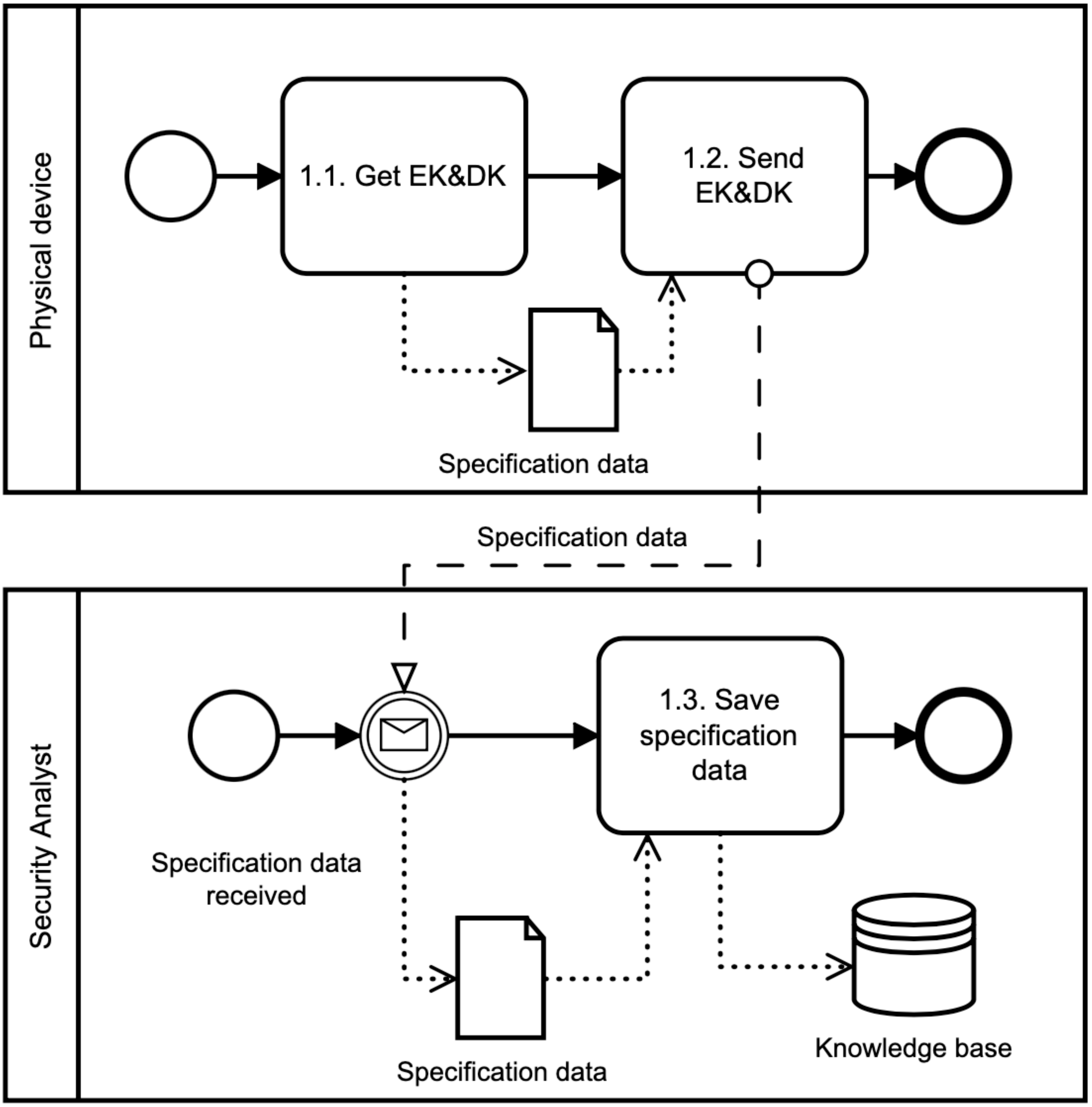}
            \caption[]%
            {{\small Save specification data.}} 
            \label{fig:pdig1}
        \end{subfigure}
        \vskip\baselineskip
        \begin{subfigure}[b]{0.685\textwidth}   
            \centering 
         \includegraphics[width=\textwidth]{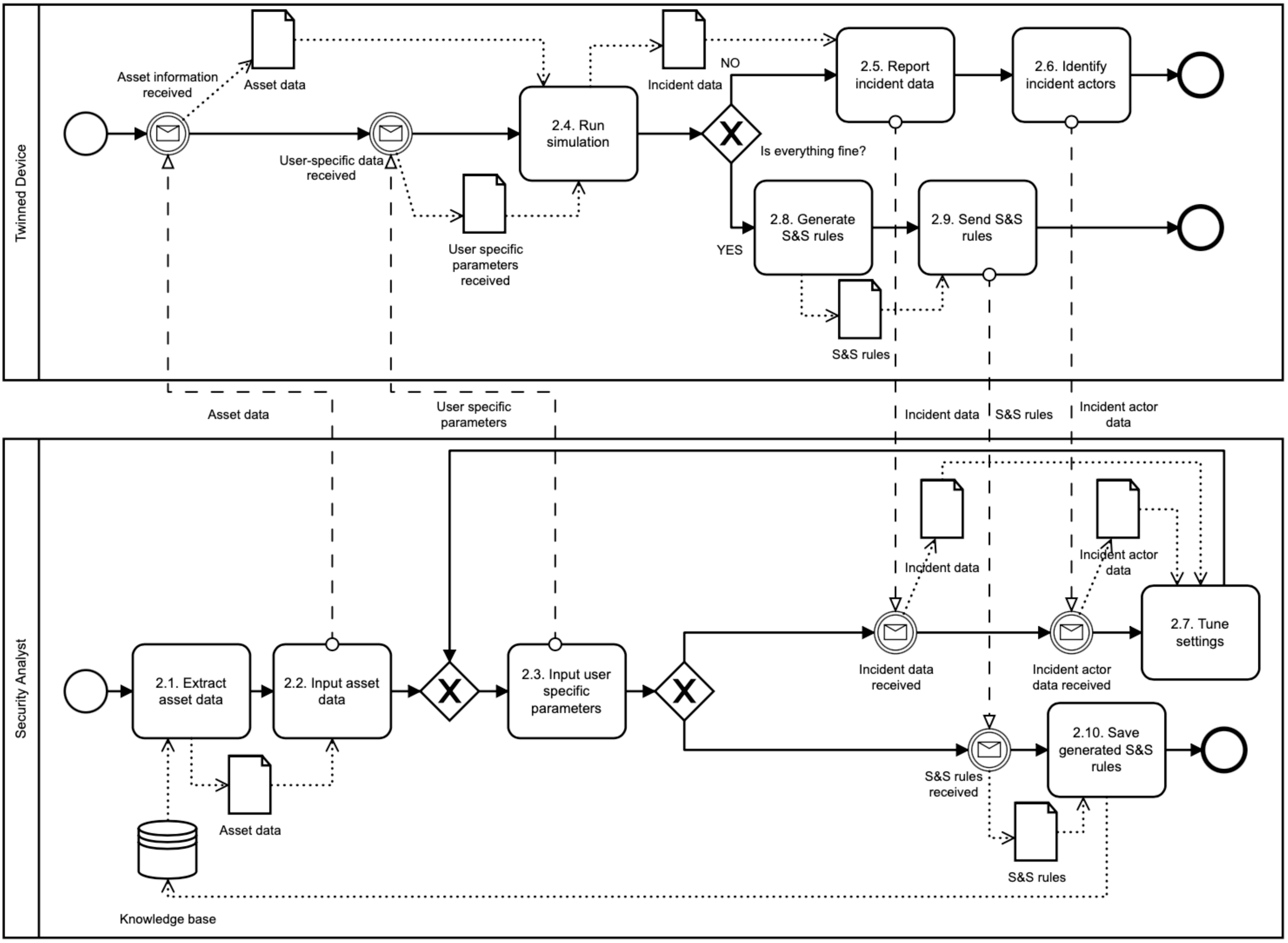}
            \caption[]%
            {{\small Generate S\&S rules.}}    
            \label{fig:pdig2}
        \end{subfigure}
        \hfill
        \begin{subfigure}[b]{0.375\textwidth}   
            \centering 
         \includegraphics[width=\textwidth]{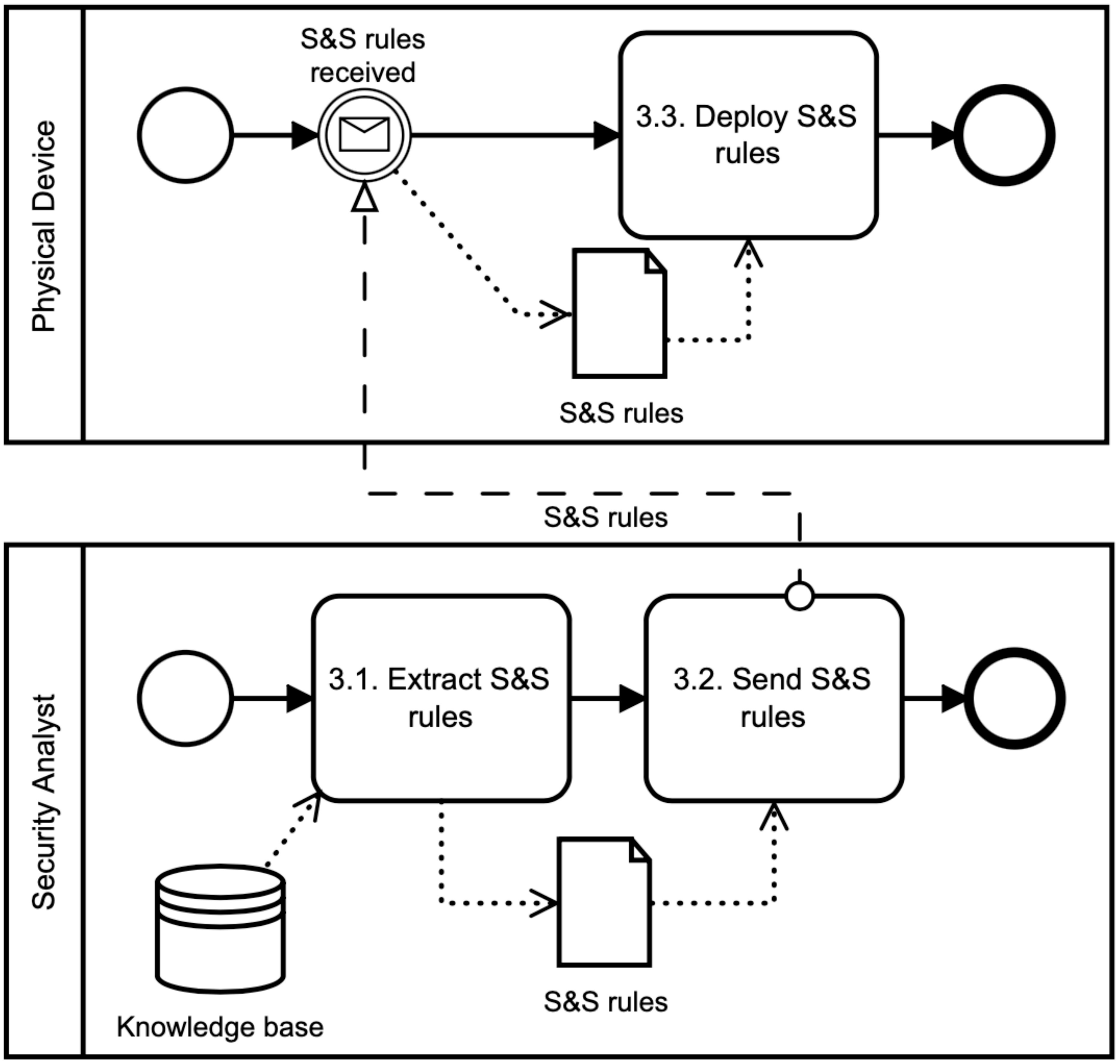}
            \caption[]%
            {{\small Deploy S\&S rules.}}    
            \label{fig:pdig3}
        \end{subfigure}
        \caption[Process-based framework. ]
        {\small Process-based framework for TTS-CPS based on digital twin security simulations.} 
        \label{fig:pd}
    \end{figure}

\subsection{Digital-physical mapping} \label{Digital-Physical}
In CPSs, as physical assets start operating, DTs run synchronously with their physical counterparts while integrating data from multiple sources, such as EK, DK, PE, to generate an abstract view of overall phenomena with the key objective to track data inconsistencies between PE and VE. Inconsistencies between the two spaces call for the adoption of better strategies that evolve DTs and physical counterparts to support accurate prediction and optimization of the underlying processes~\cite{suhail2021securing, tao2017digital}. 

To describe the process, its corresponding actors, systems and artefacts, Fig.~\ref{fig:pd} shows a process-based framework, expressed in Business Process Model and Notation (BPMN). The process consists of three key phases, see Fig.~\ref{fig:legend}, 1. Save specification data, 2. Generate S\&S rules, and 3. Deploy S\&S rules. In the first phase, see Fig.~\ref{fig:pdig1}, the security analyst receives the specification data from the physical device and saves this data to the knowledge base.
The goal of the second phase, see Fig.~\ref{fig:pdig2}, is to produce the S\&S rules. The DT runs the simulation (see task 2.4. Run simulation) using the asset data and user-specific parameters to achieve this goal. If the incident happens during the simulation, the incident data and the incident actors are submitted to the security analyst for the setting tune (see task 2.7 Tune settings). The new user-specific parameters are generated and submitted to the DT for the next simulation run. If the simulation does not generate any incident, the S\&S rules are created and saved (see task 2.10 Save generated S\&S rules) to the knowledge base. Finally, in the third phase, see Fig.~\ref{fig:pdig3}, the system analyst sends the S\&S rules to the physical device for deployment.

\subsubsection{Digital twins operation modes}
DTs operate in the three operation modes to support the comprehension of emergent system behaviour, namely replication, simulation, and data analytics.

\begin{figure}[ht!]
\centerline{\includegraphics[width=4.5in]{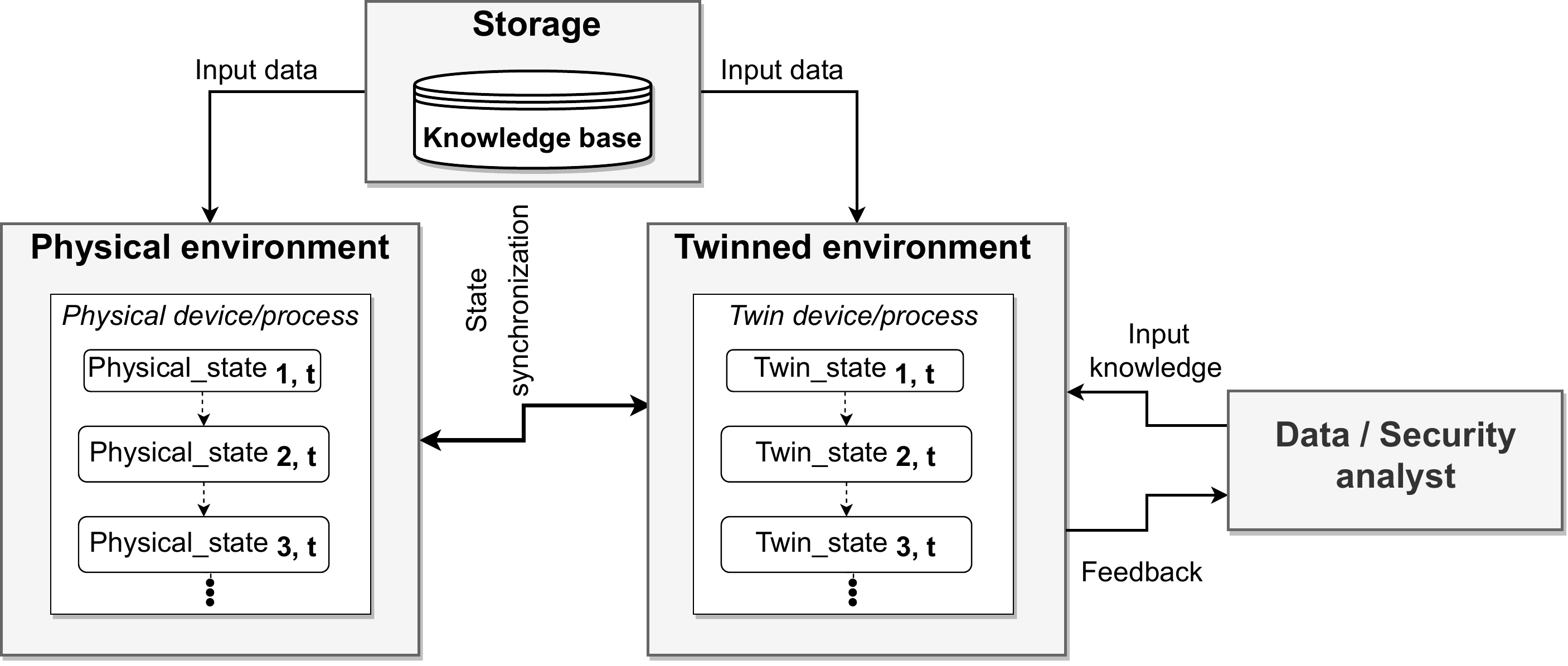}}
\caption{Replication mode of digital twin.}
\label{fig:rep_mode}
\end{figure}

For replication mode (as showcased in Fig.~\ref{fig:rep_mode}), VE and PE must be constantly connected such that VE must continuously provide digital tracing of PE events by mirroring data through log files, sensor measurements, network communication, etc. Depending on the underlying process requirements, VE data can be recorded and replay after a specific time interval or even offline. Based on closed-loop operation between the digital-physical mapping, the replication mode can provide testing and training platform where system can be trained on how to respond against advanced stealthy attacks, and defensive strategies can be tested before transmitting to real-world systems, for instance, red-blue team exercises for cybersecurity training opportunities and cyber ranges~\cite{becue2018cyberfactory}.

\begin{figure}[ht!]
\centerline{\includegraphics[width=3.0in]{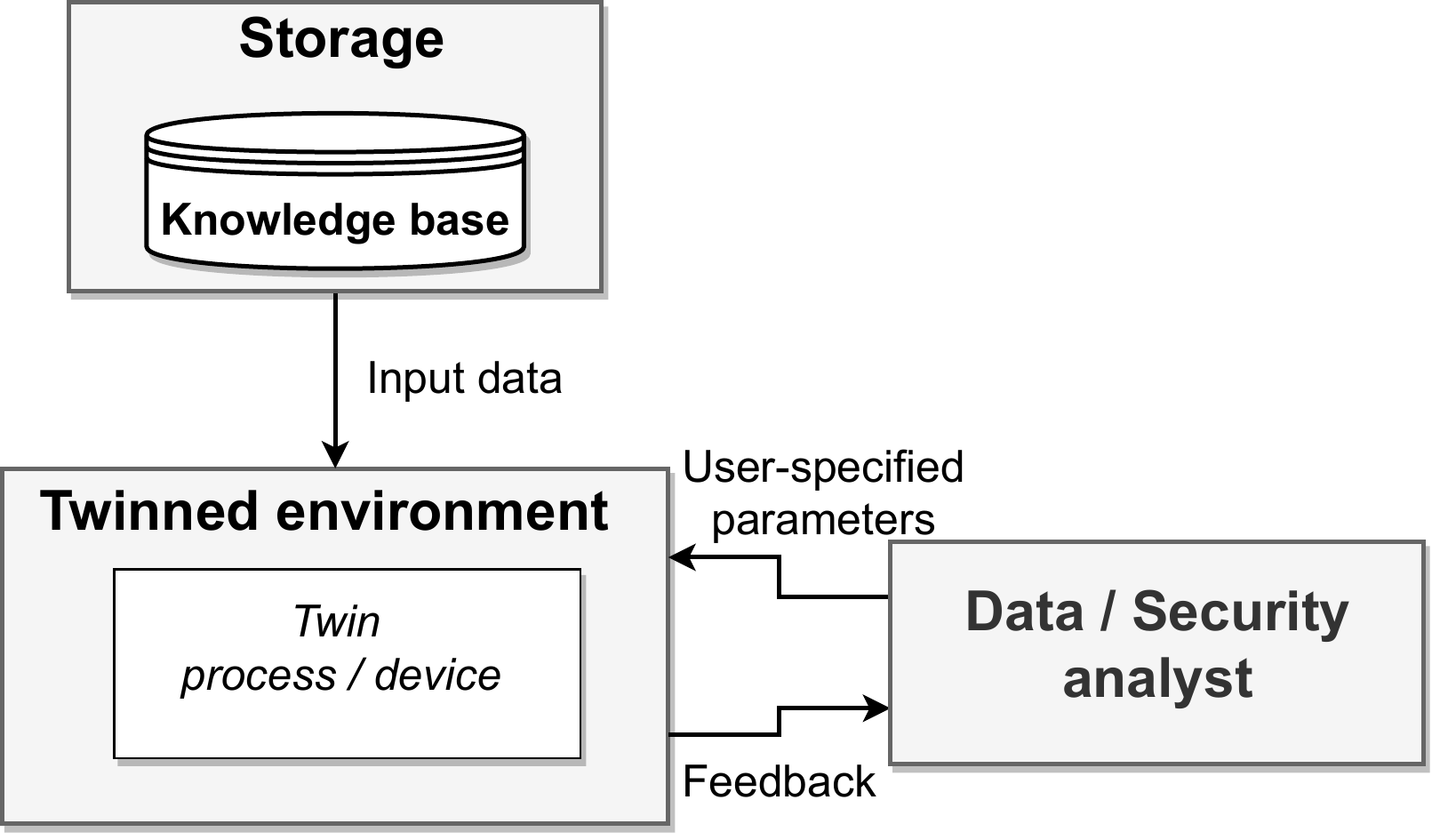}}
\caption{Simulation mode of digital twin.}
\label{fig:sim_mode}
\end{figure}

The simulation mode (as showcased in Fig.~\ref{fig:sim_mode}) runs independently of its physical counterparts by allowing running tests repeatedly by resetting the model under a broad range of specified conditions. While supporting~{\it security by design} approach, simulation mode can perform security tests within the virtual environment to analyze process changes, test devices, or detect misconfigurations, predict the possibility of attacks or system malfunctioning to carry out what-if analysis. 

Data analytics and optimization use asset's behavioural data, sensor data, and system current state or history data as valuable input to extract actionable insights while utilizing the predictive capability of machine learning algorithms available through threat intelligence. Threat intelligence can analyze the massive volume of data in real-time, learn useful patterns, and detect the presence of vulnerabilities, threat actors or inadvertent disruptions in the system, and can trigger the appropriate defence mechanisms autonomously to minimize the threat landscape~\cite{suhail2021securing}. 

For instance, threat intelligence can be integrated into the Security Information and Event Management (SIEM)~\cite{dietz2020integrating} to check the adherence of S\&S rules. Note that we only focus on the simulation mode of specification-based DT in the current work.

\subsubsection{Scenario specification}\label{scenario}
\begin{figure}[ht!]
\centerline{\includegraphics[width=4.5in]{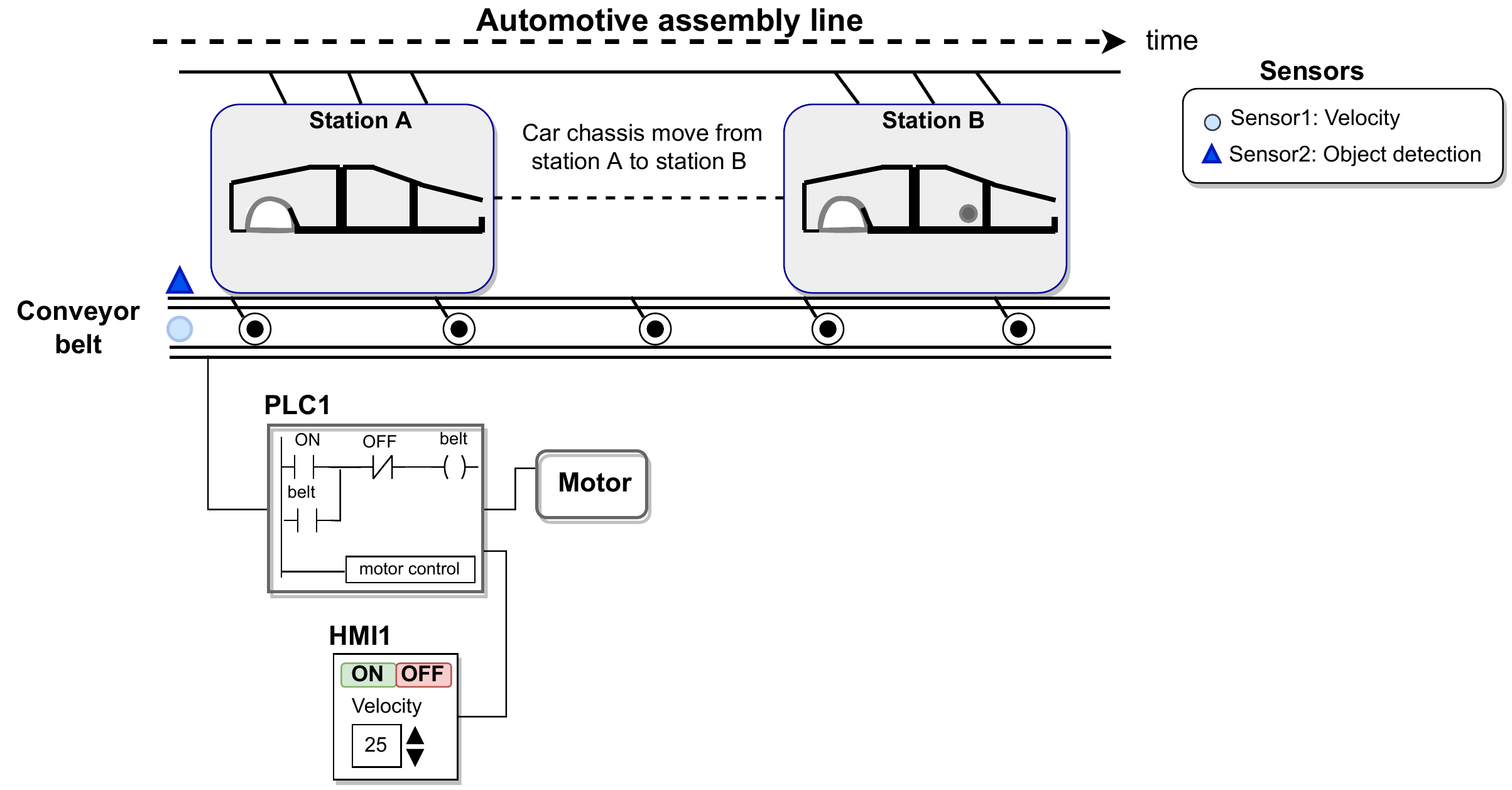}}
\caption{Scenario specification of assembly line in the automotive industry.}
\label{fig:scenario1}
\end{figure}

\begin{figure}[ht!]
\centerline{\includegraphics[width=4.5in]{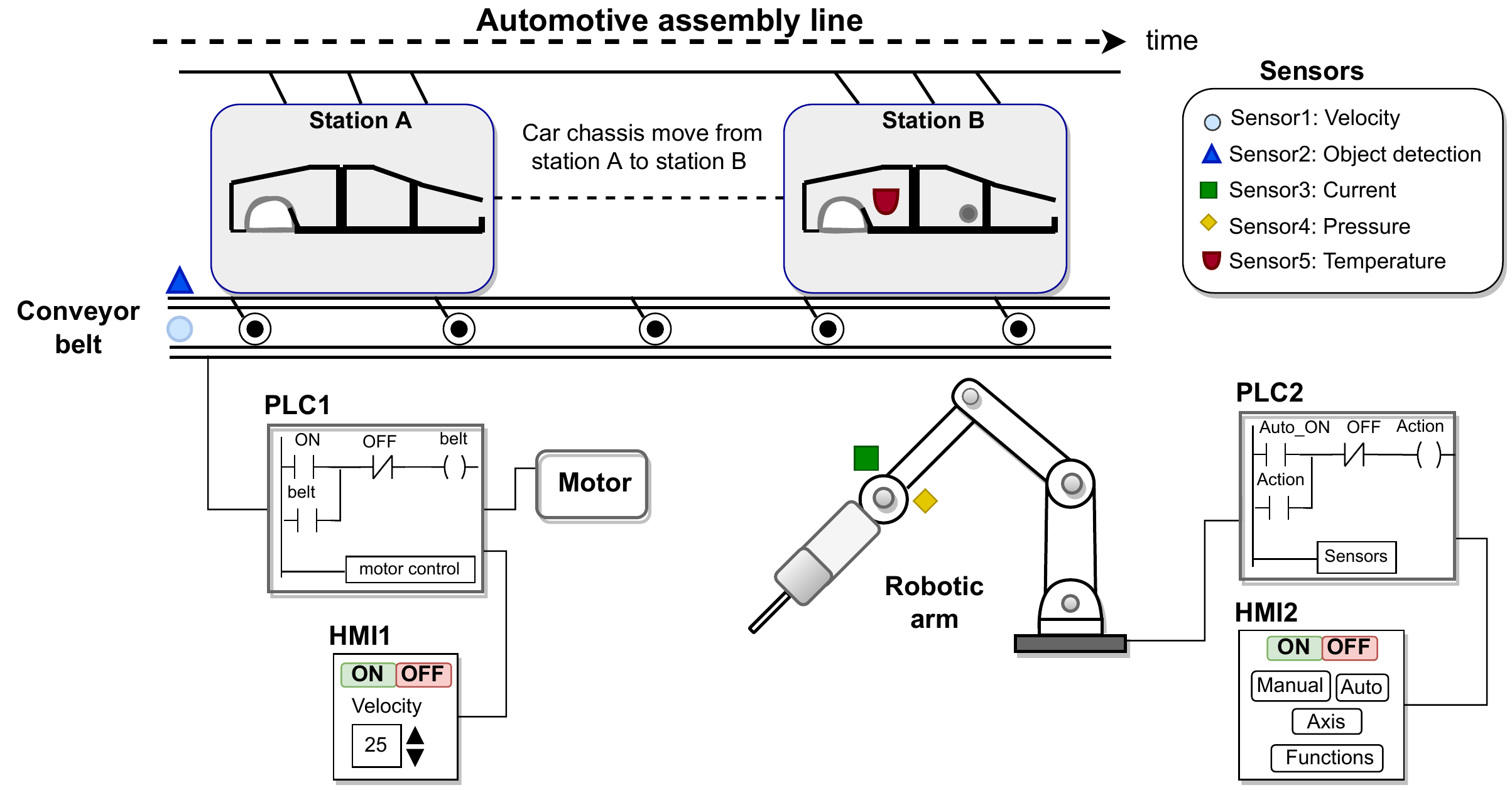}}
\caption{Scenario specification of robotic arm at the assembly line in the automotive industry. (Adapted from~\cite{suhail2021securing}).}
\label{fig:scenario2}
\end{figure}

In the following, we demonstrate the use case scenario of the automotive industry. For the sake of simplicity, we divide it in two sub-scenarios showcased in Fig.~\ref{fig:scenario1} and Fig.~\ref{fig:scenario2}.

Fig.~\ref{fig:scenario1} illustrates the exemplary physical process where we consider an assembly line in the automotive industry. The assembly line has multiple stations equipped with machinery for performing dedicated tasks. The motor-driven conveyor system moves the objects (chassis) from Station A (chassis loading point) to Station B (chassis welding point) for performing a physical task (such as spot welding). Initially, the motor is off. Firstly, the chassis are required to be loaded on the belt (as shown in Fig.~\ref{fig:scenario1}) at Station A and their presence is being detected by a proximity sensor (Sensor2: Object Detection). To do so, the velocity (Sensor1: Velocity) of the conveyor system must be monitored against a certain threshold based on the following conditions: (i) detect and load only a specific number of chassis on the belt, and (ii) maintain a safe distance to avoid collision between two adjacent chassis on the assembly line. Based on a pre-defined task duration, the chassis at Station A can be moved to Station B while more chassis can be loaded at Station A.  

Secondly, the welding operation on the chassis is performed by a robotic arm at Station B (as shown in Fig.~\ref{fig:scenario2}). To do so, the welding gun applies appropriate current and pressure at the welding spot measured through Sensor3: Current and Sensor4: Pressure. Another sensor (Sensor5: Temperature) measures the temperature during the welding process and is bounded by a threshold to avoid material deterioration. To monitor tool wear data, the robotic arm is equipped with sensors (such as vibration, force cutting, acoustic emission). During manufacturing processes, machine unavailability (equipment deterioration or machine malfunctioning) and uncertain disturbances (due to urgent job arrival or job tardiness) usually occur, leading to performance and production disruption. Therefore, we keep monitoring tool wear data of robotic arms operating in the service station of the assembly line. Recording such data ensures continual assembly line operation while triggering suitable time for machine maintenance or timely rescheduling for dynamic job-shop scheduling. We may calculate tool wear data based on the robotic arm's number of objects (chassis) welded. To command and control physical processes (for instance, turning motor and robotic arm on/off, setting the velocity of the motor, or setting current and pressure of robotic arm, etc.) through conveyor belt and robotic arm, PLC1, HMI1, PLC2, HMI2 are used.

To illustrate how to achieve optimal operating conditions ($\epsilon\ssymbol{1}$) for a given process ($P_{ID}$), we discuss the following three conditions (i) how to maintain a safe distance (can be achieved through $d$) between adjacent vehicles frame on the conveyor belt and how to maintain temperature bounded by a threshold to avoid material deterioration during welding process by robotic arm?, (ii) how to enforce data consistency between PE and VE?, and (iii) how to estimate the asset capacity ($A_{\epsilon}$) for the next production process ($P^{next}_{ID}$)? in Algorithm~\ref{algo:simulation1}, Algorithm~\ref{algo:simulation2}, and Algorithm~\ref{algo:replication}. The simulation results are presented in Section~\ref{results}.

\begin{algorithm}[!ht]
\caption{Simulation mode: {\it conveyor belt}}\label{algo:simulation1}
\begin{algorithmic}[1]
\Require $V$, $\tau^\mathbf{min}_{v}$, $\tau^\mathbf{max}_{v}$, $D_{s,o}$, ${d}$, $o_{count}$ = 0, $\tau^\mathbf{min}_{o}$, $\tau^\mathbf{max}_{o}$
\Ensure $\epsilon\ssymbol{1}$
\State $V$ = 0 \Comment{motor is off.}
\State Assume $\tau^\mathbf{min}_{v}$, $\tau^\mathbf{max}_{v}$, ${d}$,  $\tau^\mathbf{min}_{o}$, $\tau^\mathbf{max}_{o}$ under $\epsilon_i$  where i $\in$ $\{1,2, \dots, n\}$
\State Set $\tau^\mathbf{min}_{v}$ $\leq$ $V$ $\leq$ $\tau^\mathbf{max}_{v}$ \Comment{Sensor1: Velocity. Input $V$ through HMI}
\Do 
\State Check $D_{s,o}$ \Comment{Sensor2: Object Detection}
\If{$D_{s,o}$ == TRUE}
\State $o_{count}$ = $o_{count}$ + 1 \Comment{count loaded object.}
\State Check $\tau^\mathbf{min}_{o}$ $\leq$ $o$ $\leq$ $\tau^\mathbf{max}_{o}$
\State Set $d$ = $t$ seconds
\EndIf 
\State $\epsilon_{i+1}$
\doWhile{($\epsilon\ssymbol{1}$==TRUE)} 
\State Call Algorithm~\ref{algo:rules}
\State Update PE settings based on $\epsilon\ssymbol{1}$ 
\end{algorithmic}
\end{algorithm}

\begin{figure}[!ht]
        \centering
        \begin{subfigure}[b]{0.475\textwidth}
            \centering
        \includegraphics[width=\textwidth]{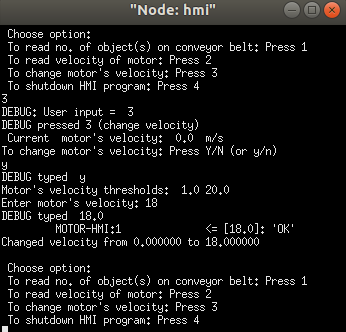}
            \caption[]%
            {{\small Human Machine Interface (HMI).}}    
            \label{fig:sim1_hmi}
        \end{subfigure}
        \hfill
        \begin{subfigure}[b]{0.475\textwidth}  
            \centering 
         \includegraphics[width=\textwidth]{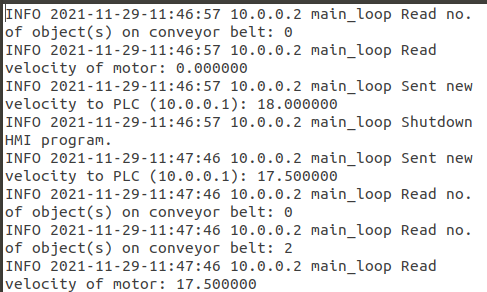}
            \caption[]%
            {{\small Excerpt of HMI log.}}    
           \label{fig:sim1_hmilog}
        \end{subfigure}
        \vskip\baselineskip
        \begin{subfigure}[b]{0.475\textwidth}   
            \centering 
         \includegraphics[width=\textwidth]{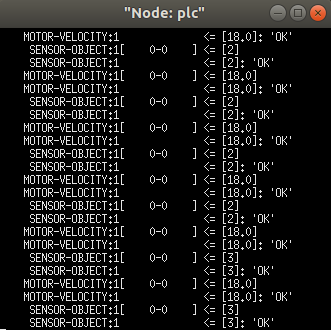}
            \caption[]%
            {{\small Programmable Logic Controller (PLC) parameters.}}    
            \label{fig:sim1_plc}
        \end{subfigure}
        \hfill
        \begin{subfigure}[b]{0.475\textwidth}   
            \centering 
         \includegraphics[width=\textwidth]{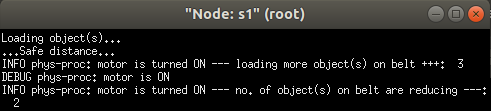}
            \caption[]%
            {{\small Physical process.}}    
           \label{fig:sim1_switch}
        \end{subfigure}
        \caption[]
        {\small Scenario I: Conveyor belt} 
        \label{fig:sim1}
    \end{figure}

In simulation mode, the current physical state of the system is not known, hence it has to rely on user-specified settings and parameters. Therefore, we assume $\epsilon_i$ values of velocity thresholds ($\tau^\mathbf{min}_{v}$, $\tau^\mathbf{max}_{v}$), delay (${d}$), and thresholds for number of objects on the belt ($\tau^\mathbf{min}_{o}$, $\tau^\mathbf{max}_{o}$) (as outlined in Algorithm~\ref{algo:simulation1}). Through $\epsilon\ssymbol{1}$, test repetitions can be realized by resetting the values and rerunning the simulation until the optimal operating conditions. Initially, the motor's velocity is 0, i.e., the conveyor belt is not operating. User can set set motor's velocity ($V$) through HMI whereas $V$ should be within the predefined thresholds  ($\tau^\mathbf{min}_{v}$, $\tau^\mathbf{max}_{v}$). Sensor1: Velocity tracks the speed of the motor. Once the belt start operating, the objects can be loaded on the belt (as shown in Fig.~\ref{fig:sim1_switch}). The objects arrival at Station A on the belt can be detected through sensor (Sensor1: Object Detection). Moreover, to minimize the makespan and to enforce production optimization we can also track the number of objects on belt through thresholds for number of objects on the belt ($\tau^\mathbf{min}_{o}$, $\tau^\mathbf{max}_{o}$). A safe distance can be maintained between two adjacent chassis on the assembly line by adjusting the velocity of the conveyor system and delay ($d$). Furthermore, S\&S can be generated or updated based on the velocity of the conveyor system that must be monitored against a certain threshold to maintain a safe distance and avoid collision.

\begin{figure}[!ht]
        \centering
        \begin{subfigure}[b]{0.475\textwidth}
            \centering
        \includegraphics[width=\textwidth]{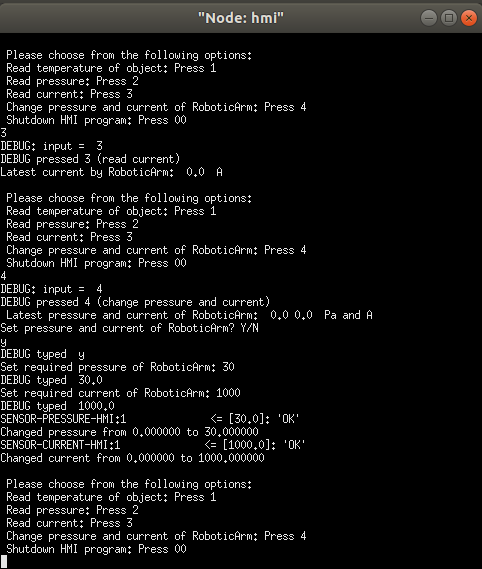}
            \caption[]%
            {{\small Human Machine Interface (HMI).}}    
            \label{fig:sim2_hmi}
        \end{subfigure}
        \hfill
        \begin{subfigure}[b]{0.475\textwidth}  
            \centering 
         \includegraphics[width=\textwidth]{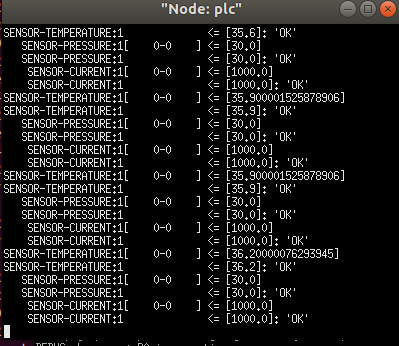}
            \caption[]%
            {{\small Programmable Logic Controller (PLC) parameters.}}    
            \label{fig:sim2_plc}
        \end{subfigure}
        \vskip\baselineskip
      \begin{subfigure}[b]{0.475\textwidth}   
            \centering 
         \includegraphics[width=\textwidth]{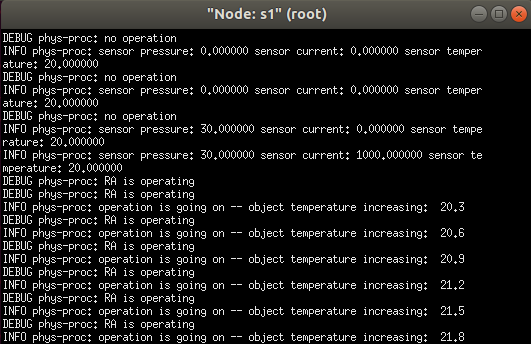}
            \caption[]%
            {{\small Physical process-a}} 
            \label{fig:sim2_switch11}
        \end{subfigure}
        \hfill
        \begin{subfigure}[b]{0.475\textwidth}   
            \centering 
         \includegraphics[width=\textwidth]{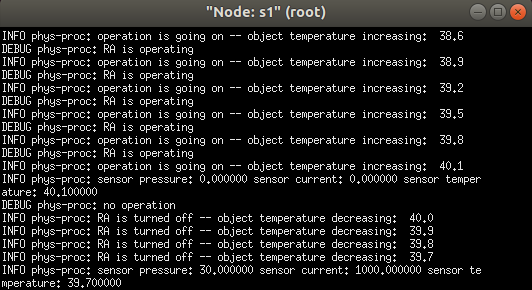}
            \caption[]%
            {{\small Physical process-b}} \label{fig:sim2_switch12}
        \end{subfigure}
        \caption[]
        {\small Scenario II: Robotic arm operating on the conveyor belt. } \label{fig:sim2}
    \end{figure}

To simulate the spot welding by robotic arm, we assume current thresholds ($\tau^\mathbf{min}_c$, $\tau^\mathbf{max}_{c}$), pressure thresholds ($\tau^\mathbf{min}_{p}$, $\tau^\mathbf{max}_{p}$), and delay ($d$). Initially the task status is set as zero. The user sets the values of current ($C$) and pressure ($P$) through HMI (as shown in Fig.~\ref{fig:sim2_hmi}) bounded by respective thresholds ($\tau^\mathbf{min}_{c}$, $\tau^\mathbf{max}_{c}$) and ($\tau^\mathbf{min}_{p}$, $\tau^\mathbf{max}_{p}$) (as outlined in Algorithm~\ref{algo:simulation2}). As the robotic arms starts welding task, current sensor ($D_\mathbf{{s,c}}$) and pressure sensor ($D_\mathbf{{s,c}}$) collect the respective data. The temperature sensor ($D_{{s,t}}$) keeps monitoring the object temperature ($o_{temp}$) to make sure it stays between predefined bounds ($\tau^\mathbf{min}_{t}$, $\tau^\mathbf{max}_{t}$) whereas to simulate the increase in $o_{temp}$ we assume value of $\Delta_{t}$ based on $C$ and $P$. For example, if the inputted values of $C$ and $P$ are closer to $\tau^\mathbf{max}_{c,p}$, the value of $\Delta_{t}$ will be higher which is ideally an appropriate assumption. 
We based our task completion ($task\_status$ == 1) on a pre-defined task duration, i.e., delay ($d$) so that the object moves to the next Station (for instance, paint shop) to undergo the next physical process. After the task completion, $o_{temp}$ begins to decrease until it reaches at room temperature ($init\_temp$) as shown in Fig.~\ref{fig:sim2_switch12}. 

\begin{algorithm}[!ht]
\caption{Simulation mode: {\it robotic arm}}\label{algo:simulation2}
\begin{algorithmic}[1]
\Require $C$, $\tau^\mathbf{min}_{c}$, $\tau^\mathbf{max}_{c}$, $D_{{s,c}}$, $P$,  $\tau^\mathbf{min}_{p}$, $\tau^\mathbf{max}_{p}$, $o_{temp}$
\Ensure $\tau^\mathbf{min}_{t}$ $\leq$ $D_{{s,t}}$ $\leq$ $\tau^\mathbf{max}_{t}$, $task\_status$==1
\State $P$ = 0, $C$ = 0, $o_{count}$ = 0, $o_{temp}$ = init\_temp, $task\_status$ == 0, \Comment{Initial conditions of the physical process}
\State Set $\tau^\mathbf{min}_{c}$ $\leq$ $C$ $\leq$ $\tau^\mathbf{max}_{c}$
\Comment{Sensor3: Current. Input $C$ through HMI}
\State Set $\tau^\mathbf{min}_{p}$ $\leq$ $P$ $\leq$ $\tau^\mathbf{max}_{p}$  \Comment{Sensor4: Pressure. Input $P$ through HMI} 
 \Do 
\State Start welding task
\State $o_{temp}$ = init\_temp + $\Delta_{t}$ \Comment{$\Delta_{t}$ is decided based on $C$ and $P$.} 
\State Check $\tau^\mathbf{min}_{t}$ $\leq$ $o_{temp}$ $\leq$ $\tau^\mathbf{max}_{t}$ \Comment{Sensor5: Temperature.}
\doWhile{($d$ = $t$ seconds)} 
\State $task\_status$ == 1
\State $o_{count}$ = $o_{count}$ + 1
\If{$o_{count}$ $\leq$ $A^\mathbf{max}_{\lambda}$} \Comment{$A_{\lambda}$ can be defined as serving 5 objects.}
\State Check equipment health 
\EndIf
\end{algorithmic}
\end{algorithm}

During the welding process, the welding gun is exposed to heat and pressure, thereby causing deformation of the welding electrodes~\cite{HONG201477}. Since maintaining equipment health defines the production quality, reduce production downtime and utility cost; therefore, we also record tool wear data. For the sake of simplicity, we based our asset capacity ($A_\lambda$) on the number of objects ($o_{count}$) being welded by the robotic arm. Upon reaching the maximum asset capacity ($A^\mathbf{max}_{\lambda}$), the equipment health should be monitored before initiating the next production process.

\begin{algorithm}[!ht]
\caption{Replication mode}\label{algo:replication}
\begin{algorithmic}[1]
\Require $\tau^\mathbf{min}_{s(k)}$, $\tau^\mathbf{max}_s(k)$
\Ensure Data consistency between PE and VE
\State Get predefined values of $\tau^\mathbf{min}_{s(k)}$ and $\tau^\mathbf{max}_{s(k)}$ from C\&V
\State Get $A^{c}_{\omega}$ of $A_{ID}$ from PE
\State Get $D_{s(k),t}$ of $S_{ID}$ from PE
\Do
\State CC = ($\tau^\mathbf{min}_{s(k)} \leq D_{s(k),t} \leq \tau^\mathbf{max}_{s(k)}$) AND ($A^{c}_{\omega}$ == on ) 
\If{(CC!=TRUE)}
\State Invoke S\&S
\If{($\mathcal{R}$!=S\&S)}
\State Call \textit{scheduling service} \Comment{Check for fault diagnosis in the equipment or the affixed sensor}
\Else
\State Call \textit{process calibration service} \Comment{Calibrate the process settings at VE}
\EndIf
\State Call Algorithm~\ref{algo:rules}
\EndIf
\doWhile{(CC==TRUE)} 
\end{algorithmic}
\end{algorithm}

In replication mode, the input knowledge (for example, actions and events in PE) are required to reproduce the same stimuli in the VE. Therefore, the real-time sensor data ($D_{s(k),t}$) and asset current state ($A^{c}_{\omega}$) from PE and threshold values ($\tau^\mathbf{min}_{s(k)}$ and $\tau^\mathbf{max}_{s(k)}$) from ICMs are recorded. The recorded events are then replayed in VE. Given that the PE is
mirrored to VE by its configuration settings, specification, and current events, the DT’s replication
mode should deliver the same results. To do so, data consistency checks are induced between the PE and the VE as outlined in Algorithm~\ref{algo:replication}. The consistency checks (CCs), for example, speed-variable CC, can be harnessed for a variety of security monitoring and operations purposes. If any inconsistent event is encountered, for instance, $D_{s(k),t}$ exceeds lower or upper bounds, the CC fails to meet the defined operational behaviour of the system. Since we pre-define S\&S rules for our situational aware CPS framework, therefore, the appropriate rules can be triggered to deal with such abnormal situations. However, in the course of advanced stealthy attacks, S\&S rules might be limited to detecting known misbehaviour. Therefore, under such circumstances, depending on attack intensity, the scheduling service is called to inspect device or network log data for fault diagnosis or process calibration service is called to reconfigure the settings. In the worst-case scenario, the affected device or service can be switched off to avoid further loss. The suggested measures are tested and verified first at the VE and afterwards regulated on the PE. Note that, the degree of state replication accuracy between PE and VE depends on the trade-off between budget and fidelity~\cite{bitton2018deriving}. S\&S can be generated or updated based on the new incident data.

\subsection{Data storage: blockchain-based digital twins} \label{BC_DTs}
DTs control and program the lifecycle of physical assets to support product servitization to end-users~\cite{suhail2021securing}. However, such advantages of DT are based on an assumption about data trust, security, and integrity~\cite{suhail2021securing}. Data trustworthiness matters more for safety-critical systems where slight dysfunction due to erroneous data may lead to wrong decisions that could imply loss of life or economic disaster. Maliciously or mistakenly, in real-life scenarios, data breaches could occur due to several reasons~\cite{suhail2021blockchainbased}. Therefore, mining actionable insights from the collected data demands a data storage infrastructure to disseminate reliable and secure data~\cite{suhail2021trustworthy}. In this regard, provenance-enabled blockchain-based DTs can be used to ensure trustworthy DTs throughout the product lifecycle~\cite{suhail2021trustworthy}.

While ensuring efficient data retrieval, the next question is what should be stored on the blockchain. Data-driven CPS primarily relies on the critical data and data-generating sources that can facilitate track and trace solutions, in addition to user- or application-specific requirements. By following~\cite{suhail2021trustworthy}, in TTS-CPS, we limit the time-consuming frequent access to the blockchain-based storage system by explicitly separating the less dynamic (or static) data and the real-time dynamic data. For instance, data from the ICMs can be considered less dynamic data as it infrequently changes with time along the lifecycle of the physical counterpart, such as provenance data, device configuration settings, system's historical data, policies and access levels. Similarly, to strengthen the rationale for integrating blockchain with DT, S\&S rules must be stored and retrieved from the ledger (as shown in Fig.~\ref{fig:framework2}) to ensure their reliability. In essence, blockchain inherently retains the history of modifications and thus can circumvent illegal data modification that may lead to other data-related problems.

Being the virtual replicas of their physical counterparts, the DTs shares the operational behaviour of the underlying physical process or device~\cite{Eckhart2019, suhail2021securing}. On the flip side, the attackers may exploit the valuable knowledge about the system accessible through DTs to put DTs into a malicious state~\cite{suhail2021blockchainbased, suhail2022security}. Thus to avoid DTs use case as abuse case, DTs can be audited (for instance, changing the simulation setup parameter or state data) by using provenance-aware blockchain-based solutions. 

\section{Implementation and Evaluation}\label{simulation}
In order to evaluate the proposed framework, a simulation for a DT of an assembly line is implemented. The proof of concept demonstrates that the virtual environment conforms to the defined ICMs. Additionally, for building the process knowledge of DTs, acquiring reliable system specification data from blockchain establishes confidence in the DTs, thereby avoiding DT abuse cases. In the following, we demonstrate the viability of the proposed framework in a proof of concept, including the generation of DTs and the formal verification. 

\subsection{Simulation and results}\label{results}
For our use case scenario showcased in Fig.~\ref{fig:scenario1} and Fig~\ref{fig:scenario2}, we use MiniCPS~\cite{antonioli2015minicps} which is built on Mininet~\cite{lantz2010network}-a prototyping environment for networks. MiniCPS extents Mininet to emulate and simulate CPS network components, including PLCs, HMIs, and industrial network communication over EtherNet/IP or
Modbus/TCP. We based our work on the prototype available on GitHub~\footnote{\url{https://github.com/FrauThes/DigitalTwin-ConveyorBelt}}. To create the virtual environment (discussed in Section~\ref{scenario}), the main input is obtained from the engineering- and domain-knowledge (Section~\ref{subsec:EK} and Section~\ref{subsec:DK} respectively.) The initial simulation setup must conform to the defined rules (Section~\ref{subsec:rules}). The system events are logged to a file. For example, Fig.~\ref{fig:sim1_hmilog} shows excerpt of HMI log. Similarly, log files are maintained for PLCs and other processes. The purpose of logged statements is to alert plant operators or security professionals about the system's abnormal condition to carry out the investigation by tracking and tracing the entities. Additionally, the log files acquired from multiple instances of simulation can be scrutinize to gain more insights into the underlying events. The log files can be used to generate or update rules upon the execution of simulation. The logged events can be further utilized by Security Information and Event Management (SIEM)~\cite{dietz2020integrating} and/or threat hunting~\cite{suhail2021securing}.

We have used Hyperledger Fabric- a permissioned blockchain to store and retrieve ICMs and other information significant to the process. The simulation mode runs independently of its physical counterpart and provide~{\it trial and error} approach~\cite{suhail2021securing} while monitoring the states or events of the physical process. Therefore, specification data can be retrieved and stored before and after the simulation, respectively. After repeatedly resetting the model through a broad range of specified conditions, the obtained results can be used to update the twin and eventually the physical asset. Fig.~\ref{fig:sim1} showcases a conveyor belt scenario, whereas Fig.~\ref{fig:sim2} showcases a robotic arm operating on a conveyor belt. The simulation setup is implemented based on the specification data (such as ICMs). 

\subsection{Formal verification of the TTS-CPS}
This section presents the formal verification of our blockchain-based DT model. In the verification process, we demonstrate the correctness of the base system. We need system specification and properties to verify a proposed model or a system~\cite{malik2013modeling}. We have used bounded model checking to evaluate the correctness of the underlying properties. The simulation model of conveyor belt and robotic arm (as discussed in Section~\ref{scenario}) is first translated into Satisfiability Modulo Theories Library (SMT-Lib) and then the Z3 Solver is used to perform the verification. More details about SMT-Lib and Z3 Solver can be obtained from~\cite{malik2013modeling} and~\cite{malik2015modeling}. In bounded model checking, the goal is to evaluate the correctness of the system inputs that drives the system into a state where the system always terminates after a finite number of steps. Formally, bounded model checking is defined as a Kripke Structure and a bound $k$, where the problem is to find ${M \models _k Ef}$. In bounded model checking problem, an execution path is tried to be searched in a Kripke structure $M$ of length $k$ that satisfies a formula $f$. We have verified the blockchain-based DT framework by proving the correctness of conveyor belt and robotic arm stated in Algorithm~\ref{algo:simulation1} and Algorithm~\ref{algo:simulation2} respectively. We have translated the aforesaid algorithms into SMT-Lib and then defined certain correctness properties to verify the algorithms using the Z3 solver. 

SMT has their roots in Boolean Satisfiability (SAT) Solvers. It is generally used and is a part of automated deduction in for satisfiability of formulas over some theories on interests~\cite{de2009satisfiability}. A common benchmark framework and input platform is provided by the SMT that usually helps in evaluating the systems~\cite{smtlib}. The SAT and SMT solver performs differently in a way that SAT evaluates the satisfiability of propositional formulas. On the other hand, SMT performs the satisfiability of first-order logic formulas based on underlying theories.~\cite{frade2011verification}. Deductive verification is one of the many fields in which SMT has been used. Considering the sensitive nature of recent computer sciences applications, which involves modeling and planning, performing formal analysis and verification through SMT is considered an important task.~\cite{smtlib}. Some examples are available in~\cite{malik2012methodology} and~\cite{malik2012convergence}. Several solvers supports the implementation of SMT-lib. Some examples includes NuSMV, CVC4, OpenSMT, and SathSAT5. The classification of the solvers can be done based on the underlying theories, logic, and interface~\cite{frade2011verification}. In our study, we have used a theorem prover, namely Z3, which is developed by Microsoft for checking automated satisfiability. It evaluates if the model satisfies the properties specified in SMT-lib. More information regarding the use and application of Z3 can be found in~\cite{de2008z3}.

\begin{defi}[Bounded Model Checking~\cite{biere2003bounded}]
Formally, given a Kripke Structure $(S,S_o,R,L)$ and a $k$ bound, the bounded model checking problem is to find $M\models_{k}Ef$ where: $S$ is the finite set of states, $S_o$ is a set of initial states, $R$ is the set of transitions, such that $R\subseteq S\times SL$ is the set of labels. \end{defi}

The bounded model checking problem is to find an execution path in $M$ of length $k$ that satisfies a formula $f$. Kripke structure, which is a state transition graph, is used to represent the behaviour of the system~\cite{malik2013modeling}. In Kripke structure nodes are the set of reachable states of the system, edge represents the transitions, and label functions map nodes to the set of properties hold in the state. A path in a Kripke structure can be stated as an infinite sequence of states represented $\rho = S_1,S_2,S_3 \dots$ such that for $\forall i\geq 0, (S_i, S_{i+1})\in R$. The model $M$ may produce a path set = $S_1,S_2,S_3 \dots$. To describe the property of a model some formal language, such as Computation Tree Logic (CTL*), CTL, or Linear Temporal Logic (LTL) can be used.

\begin{defi}[SMT Solver~\cite{cordeiro2011smt}]
Given a theory $\beth$ and a formula $f$, the SMT solvers perform a check whether $f$ satisfies $\beth$ or not. 
\end{defi}

To perform the verification of the models using Z3 (an SMT Solver), we unroll the model $M$ and the formula $f$ that provides $M_k$ and $f_k$, respectively. Moreover, the said parameters are then passed to Z3 to check if $M_k \models_\Gamma f_k$~\cite{smtlib}. The solver will perform the verification and provide the results as satisfiable (sat) or unsatisfiable (unsat). If the answer is~\emph{sat}, then the solver will generate a counterexample, which depicts the violation of the property or formula $f$. Moreover, if the answer is~\emph{unsat}, then the formula or the property $f$ holds in $M$ up to the bound $k$ (in our case, $k$ is execution time).

We have identified certain properties, which we have verified in the conveyor belt and robotic arm algorithm. The properties are as follows.

\begin{Properties}
\item (Time) ``Does the welding performed by the robotic arm follows the specified timeline, i.e., $d=t$''? 
\item (Temperature) ``Does the robotic arm violates the welding temperature range, i.e., $\tau^\mathbf{min}_t$ $\leq$ $o_{temp}$ $\leq$ $\tau^\mathbf{max}_t$''?  
\item (Velocity) ``Does the conveyor belt moves according to the specified range, i.e., $\tau^\mathbf{min}_v$ $\leq$ $V$ $\leq$ $\tau^\mathbf{max}_v$''?
\end{Properties}

\begin{figure}[ht!]
\centerline{\includegraphics[width=3.5in]{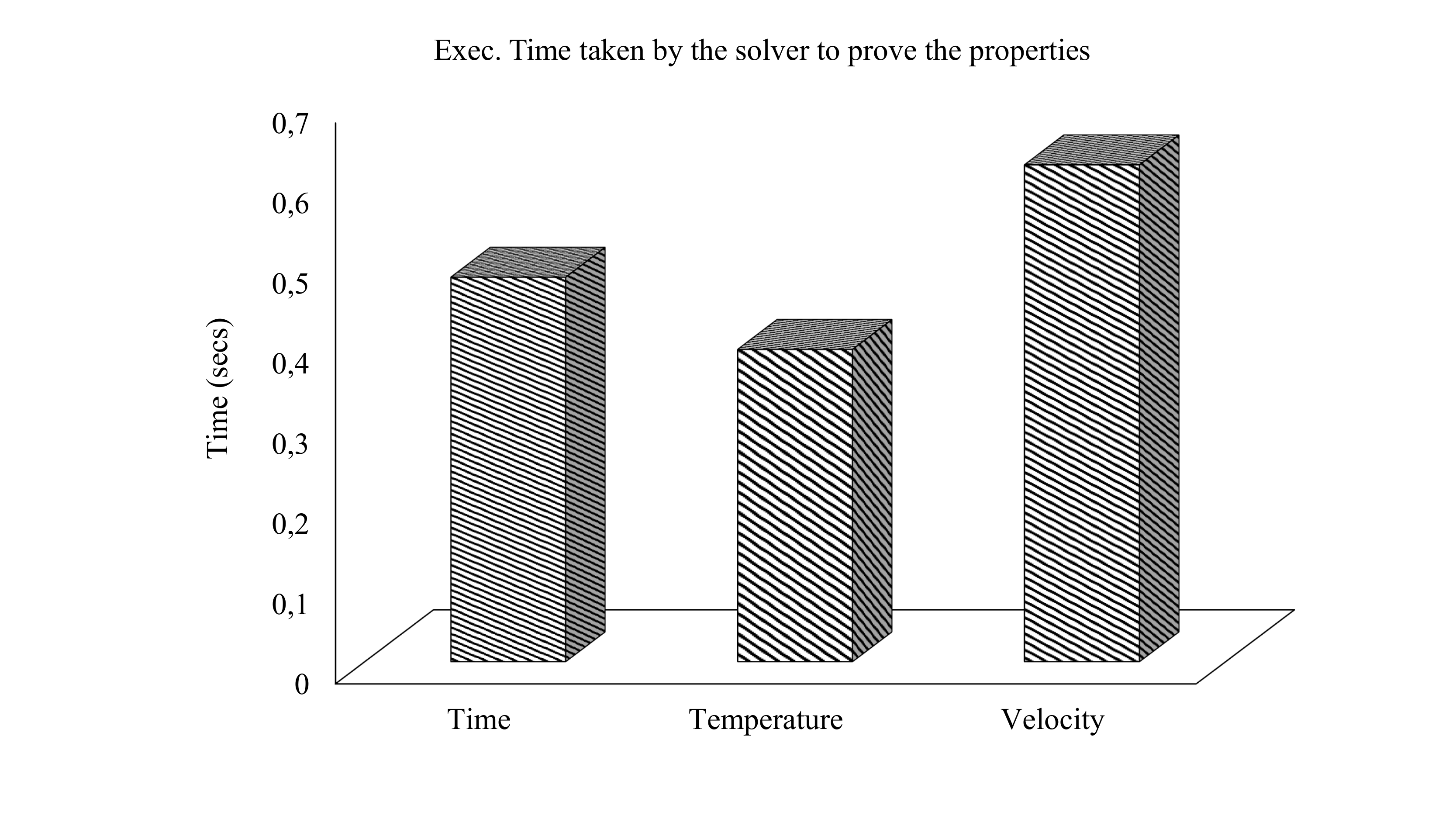}}
\caption{Verification results of robotic arm and conveyor belt algorithms.}
\label{fig:properties}
\end{figure}

The verification results of the properties are shown in Fig.~\ref{fig:properties}. The time on y-axis represents the execution time taken by the solver to verify each property. As stated above, if the properties are not violated, the solver generates~\enquote{unsat}, which depicts that the solver was unable to find any sequence of executions within the models that violate the stated properties. 

\section{Conclusion and Outstanding Challenges}
\label{conclusion}
In this work, we have targeted two critical challenges, i.e., (i) how to establish a situational-aware and secure CPS through DTs and (ii) how to establish the trustworthy generation of specification-based DTs. To address these issues, we have proposed a TTS-CPS framework. We demonstrate the feasibility of the TTS-CPS framework for an assembly line in the automotive industry through a prototypical implementation supporting the simulated network topology, PLCs, HMIs, and physical devices. Moreover, we perform formal modelling and verification of the TTS-CPS. 

An interesting direction for future work is to investigate how to construct a fault-tolerant system. In other words, in the course of undesirable incidents, it is essential to improve the system resilience during which the system enters a fail-safe state and maintain an adequate control of the physical process. Another open direction is to extend our work to carry out localization of attack, i.e., upon anomaly detection, finding the root cause of the deviation and localizing the compromised node (sensor or actuator). Furthermore, we plan to incorporate smart contracts to automate event-based processes such as triggering the appropriate defence mechanisms through S\&S rules and modifying simulation setup parameters~\cite{suhail2021blockchainbased}. Such automation ensures the benign behaviour of DTs, particularly during the replication mode due to cyclic updates.

In general, several other technical challenges need to be addressed, for instance, data trustworthiness, particularly in hierarchical DTs or a combination of DT instances that mimic the bigger picture of the CPS.
Such issues become more challenging due to data fusion from multimodal systems and uncertain scenarios due to the dynamism and complexity of underlying (sub) systems. Other issues such as data storage and performance implications stemming from blockchain and DTs' integration also need due attention.~\cite{suhail2021blockchainbased} provides a detailed discussion on the challenges that impede the successful implementation of blockchain-based DTs in the industry.

\end{document}